\documentclass{llncs}
\usepackage{psfig}
\begin{document}

\date{}
\title{Learning from the Success of MPI}
\author{William D. Gropp} %
\institute{Mathematics and Computer Science Division,
  Argonne National Laboratory,
  Argonne, Illinois 60439, \email{gropp@mcs.anl.gov},\\
  WWW home page: \texttt{www.mcs.anl.gov/\homedir gropp}}
\maketitle
\begin{abstract}
The Message Passing Interface (MPI) has been extremely successful as a
portable way to program high-performance parallel computers.  This success has
occurred in spite of the view of many that message passing is
difficult and that other approaches, including automatic
parallelization and directive-based parallelism, are easier to use.
This paper argues that MPI has succeeded because it addresses
\emph{all} of the important issues in providing a parallel programming
model.  
\end{abstract}

\section{Introduction}
The Message Passing Interface (MPI) is a very successful approach for
writing parallel programs.  Implementations of MPI exist for most
parallel computers, and many applications are now using MPI as the way
to express parallelism (see \cite{mpi-biblist-web} for a list
of papers describing applications that use MPI).  The reasons for the success
of MPI are not 
obvious.  In fact, many users and researchers complain about the
difficulty of using MPI.  Commonly raised issues include the
complexity of MPI (often as measured by the number of functions),
performance issues (particularly the latency or cost of communicating
short messages), and the lack of compile or runtime help (e.g., compiler
transformations for performance; integration with the underlying
language to simplify the handling of arrays, structures, and native
datatypes; and debugging).  More subtle issues, such as the complexity
of nonblocking communication and the lack of elegance relative to a
parallel programming language, are also raised \cite{Hansen:1998:EMI}.
With all of these criticisms, why has MPI enjoyed such
success?

One might claim that MPI has succeeded simply because of its
\emph{portability}, that is, the ability to run an MPI program on most
parallel platforms.  But while portability was certainly a necessary
condition, it was not sufficient.  After all, there were other,
equally portable programming models, including many message-passing
and communication-based models.  For example, the \texttt{socket}
interface was (and remains) widely available and was used as an underlying
communication layer by other parallel programming packages, such as
PVM \cite{pvmbook} and p4 \cite{p4-book}.  An obvious second
requirement is that of \emph{performance}: the ability of the
programming model to deliver the available performance of the
underlying hardware.  This clearly distinguishes MPI from interfaces
such as sockets.  However, even this is not enough.  This paper
argues that six requirements must \emph{all} be
satisfied for a parallel programming model to succeed, that is, to be widely
adopted.  Programming models that address a subset 
of these issues can be successfully applied to a subset of
applications, but such models will not reach a wide audience in
high-performance computing.    

\section{Necessary Properties}
The MPI programming model describes how separate \emph{processes}
communicate.  In MPI-1 \cite{mpi-forum:journal}, communication occurs
either through point-to-point (two-party) message passing or through
collective (multiparty) communication.  Each MPI process executes a
program in an address space that is private to that process.

\subsection{Portability}
Portability is the most important property of a programming model for
high-performance parallel computing.  The high-performance computing
community is too small to dictate solutions and, in particular, to
significantly influence the direction of commodity computing.
Further, the lifetime of an application (often ten to twenty years, rarely
less than five years) greatly exceeds the lifetime of any particularly
parallel hardware.  Hence, any application must be prepared
to run effectively on many generations of parallel computer, and that goal
is most easily achieved by using a portable programming model.

Portability, however, does not require taking a ``lowest common
denominator'' approach.  A good design allows the use of 
performance-enhancing features without mandating them.  For example,
the message-passing semantics of MPI allows for the direct copy of
data from the user's send buffer to the receive buffer without any
other copies.\footnote{This is sometimes called a zero-copy
transfer.}  However, systems that can't provide this direct copy
(because of hardware limitations or operating system restrictions) are
permitted, under the MPI model, to make one or more copies.  Thus MPI
programs remain portable while exploiting hardware capabilities.  

Unfortuately, portability does not imply portability with performance,
often called \emph{performance portability}.  Providing a way to
achieve performance while maintaining portability is the second requirement.

\subsection{Performance}
\label{sec:performance}
\begin{figure}
\centerline{\psfig{file=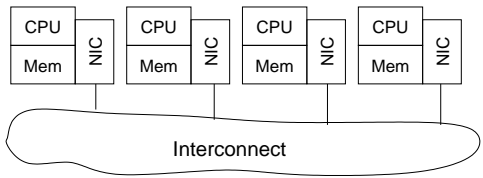,width=2.5in}}
\caption{A typical parallel computer}
\label{fig:generic-parallel}
\end{figure}

MPI enables performance of applications in two ways.  For small
numbers of processors, MPI provides an effective way to manage the use
of memory.  To understand this, consider a typical parallel computer
as shown in Figure~\ref{fig:generic-parallel}.

The memory near the CPU, whether it is a large cache (symmetric
multiprocessor) or cache and memory (cluster or NUMA), may be accessed
more rapidly than far-away memory.  Even for shared-memory computers,
the ratio of the number of cycles needed to access memory in L1 cache
and main memory is roughly a hundred; for large, more loosely
connected systems the ratio can exceed ten to one hundred thousand.
This large ratio, even between the cache and local memory, means that
applications must carefully manage memory locality if they are to
achieve high performance.  

The separate processes of the MPI programming model provide a natural
and effective match to this property of the hardware.  

This is not a new approach.  The C language provides \texttt{register},
originally intended to aid compilers in coping with a two-level memory
hierarchy (registers and main memory).  Some parallel languages, such as
HPF \cite{hpf-handbook}, UPC \cite{Carlson99}, or CoArray Fortran
\cite{Numrich:1998:CAF},  
distiguish between local and shared data.  Even programming models
that do not recognize a distinction between local and remote memory,
such as OpenMP, have implementations that often require techniques
such as ``first touch'' to ensure that operations make effective use
of cache.
The MPI 
model, based on communicating processes, each with its own
memory, is a good match to current hardware.  

For large numbers of processors, MPI also provides effective means to
develop scalable algorithms and programs.  In particular, the
collective communication and computing routines such as \texttt{MPI\_Allreduce}
provide a way to express scalable operations without exposing
system-specific features to the programmer.  
Also important for supporting scalability is the ability to express
the most powerful scalable algorithms; this is discussed in
Section~\ref{sec:modularity}. 

Another contribution to MPI's performance comes from its ability to
work with the best compilers; this is discussed in
Section~\ref{sec:composability}.  Also discussed there is how MPI
addresses the performance-tradeoffs in using threads with MPI programs.

Unfortunately, while MPI achieves both portability and performance, it
does not achieve perfect performance portability, defined as providing a
single source that runs at (near) acheivable peak performance on all
platforms.  This lack is sometimes given as a criticism of MPI, but it
is a criticism that most other programming models also share.  For
example, Dongarra et al \cite{Dongarra84c} describe six different ways to
implement matrix-matrix multiply in Fortran for a single processor;
not only is no one of the six optimal for all platforms but \emph{none} of the
six are optimal on modern cache-based systems.  Another example is the
very existence of vendor-optimized implementations of the Basic
Linear Algebra Subroutines (BLAS).  These are functionally simple and
have implementations in Fortran and C; if compilers (good as they are)
were capable of 
producing optimal code for these relatively simple routines, the
hand-tuned (or machined-tuned \cite{Whaley:2001:AEO}) versions would
not be necessary. 
Thus, while
performance portability is a desirable goal, it is unreasonable to
expect parallel programming models to provide it when uniprocessor
models cannot.
This difficulty also explains why relying on
compiler-discovered parallelism has usually failed: the problem
remains too difficult.  Thus a successful programming model must allow
the programmer to help.

\subsection{Simplicity and Symmetry}
The MPI model is often criticized as being large and complex, based on
the number of routines (128 in MPI-1 with another 194 in
MPI-2).  The number of routines is not a relevant measure, however.
Fortran, for example, has a large number of intrinsic functions; C and
Java rely on a large suite of library routines to achieve external
effects such as I/O and graphics; and common development frameworks
have hundreds to thousands of methods.

A better measure of complexity is the number of concepts that the user
must learn, along with the number of exceptions and special cases.
Measured in these terms, MPI is actually very simple.

Using MPI requires learning only a few concepts.  Many MPI programs
can be written with only a few routines; several subsets of
routines are commonly recommended, including ones with as few as
six functions.  Note the plural: for different purposes, different
subsets of MPI are used.  For example, some
recommend using only collective
communiation routines; others recommend only a few of the
point-to-point routines.  One key to the success of MPI is that these
subsets can be used without learning the rest of MPI; in this sense,
MPI is simple.  Note that a smaller set of routines would \emph{not}
have provided this simplicity because, while some applications would
find the routines that they needed, others would not.

Another sign of the effective design in MPI is the use of a single
concept to solve multiple problems.  This reduces both the number of
items that a user must learn and the complexity of the
implementation.  For example, the MPI communicator both describes the
group of communicating processes and provides a separate communication
context that supports component-oriented software, described in more
detail in Section~\ref{sec:modularity}.  Another example is
the MPI datatype; datatypes describe both the type (e.g., integer,
real, or character) and layout (e.g., contiguous, strided, or indexed)
of data.  The MPI datatype solves the two problems of describing the
types of data to allow for communication between systems with
different data representations and of describing noncontiguous data
layouts to allow an MPI implementation to implement zero-copy data
transfers of noncontiguous data.

MPI also followed the principle of \emph{symmetry}: wherever
possible, routines were added to eliminate any exceptions.  An
example is the routine \texttt{MPI\_Issend}.  MPI provides a
number of different send modes that correspond to different,
well-established communication approaches.  Three of these modes are
the regular send (\texttt{MPI\_Send}) and its nonblocking versions
(\texttt{MPI\_Isend}), and the synchronous send 
(\texttt{MPI\_Ssend}).  To maintain symmetry, MPI also provides the
nonblocking synchronous send \texttt{MPI\_Issend}.  This send mode
is meaningful (see \cite[Section 7.6.1]{gropp-lusk-skjellum:using-mpi2nd})
but is rarely used. 
Eliminating it would have removed a routine, slightly simplifying the
MPI documentation and implementation.  It would have created
an exception, however.  Instead of each MPI send mode having a nonblocking
version, only some send modes would have nonblocking versions.  Each
such exception adds to the burden on the user and adds complexity: it
is easy to forget about a routine that you never use; it is harder to
remember arbitrary decisions on what is and is not available.  

A place where MPI may have followed the principle of symmetry too far
is in the large collection of routines for manipulating groups of
processes. Particularly in MPI-1, the single routine
\texttt{MPI\_Comm\_split} is all that is needed; few users need to
manipulate groups at all.  Once a 
routine working with MPI groups was introduced, however, symmetry required
completing the set.  Another place is in canceling of sends, where
significant implementation complexity is required for an operation of
dubious use.

Of course, more can be done to simplify the use of MPI.  Some
possible approaches are discussed in Section~\ref{sec:improving-mpi}.

\subsection{Modularity}
\label{sec:modularity}

Component-oriented software is becoming increasingly important.  In
commecial software, software components implementing a particular
function are used to implement a clean, maintainable service.  In
high-performance computing, components are less common, with many
applications being built as a monolithic code.  However, as
computational algorithms become more complex, the need to exploit
software components embodying these algorithms increases.  

For example, many modern numerical algorithms for the solution of partial
differential equations are hierarchical, exploiting the structure of
the underlying solution to provide a superior and scalable solution algorithm.
Each level in that hierarchy may require a different solution
algorithm; it is not unusual to have each level require a different
decomposition of processes. 
Other examples are intelligent design
automation programs that run application components such as fluid
solvers and structural analysis codes under the control of a
optimization algorithm.  

MPI supports component-oriented software.  Both describe the subset
of processes participating in a component and to ensure that all MPI
communication is kept within the component, MPI introduced the
\emph{communicator}.\footnote{The context part of the communicator was
inspired by Zipcode \cite{skjellum.smith.ea.93}.}  Without something like a
communicator, it is 
possible for a message sent by one component and intended for that
component to be received by another component or by user code.
MPI made reliable libraries possible.

Supporting modularity also means that certain powerful variable layout
tricks (such as assuming that the variable \texttt{a} in an SPMD program
is at the same address on all processors) must be modified to
handle the case where each process may have a different stack-use
history and variables may be dynamically allocated with different base
addresses.  Some programming models have assumed that all processes
have the same layout of local variables, making it difficult or
impossible to use those programming models with modern adaptive algorithms.

Modularity also deals with the complexity of MPI.  Many tools have
been built using MPI to provide the communication substrate; these
tools and libraries 
provide the kind of easy-to-use interface for domain-specific
applications that some developers feel are important; for example,
some of these tools eliminate all evidence of MPI from the user program. 
MPI makes those
tools possible.  Note that the user base of these domain-specific
codes may be too small to justify vendor-support of a parallel
programming model.  

\subsection{Composability}
\label{sec:composability}

One of the reasons for the continued success of
Unix is the ease with which new solutions can be built by composing
existing applications.  

MPI was designed to work with other tools.  This capability is vital, because 
the 
complexity of programs and hardware continues to increase.  For
example, the MPI specification was designed from the beginning to be
thread-safe, since threaded parallelism was seen by the MPI Forum as a
likely approach to systems built from a collection of SMP nodes.  MPI-2
took this feature even further, acknowledging that there are performance
tradeoffs in 
different degrees of threadedness and providing a mechanism for the
user to request a particular level of thread support from the MPI
library.  Specificically, MPI defines several degrees of thread support.
The first, called \texttt{MPI\_THREAD\_SINGLE}, specifies that there
is a single thread of execution.  This allows an MPI implementation to
avoid the use of thread-locks or other techniques necessary to ensure
correct behavior with multithreaded codes.  Another level of thread
support, \texttt{MPI\_THREAD\_FUNNELLED}, specifies that the process may
have multiple threads but all MPI calls are made by one thread.  This
matches the common use of threads for loop parallelism, such as the
most common uses of OpenMP.  A third level,
\texttt{MPI\_THREAD\_MULTIPLE}, allows multiple threads to make MPI
calls.  While these levels of thread support do introduce a small
degree of complexity, they reflect MPI's pragmatic approach to
providing a workable tool for high-performance computing.

The design of MPI as a library means that MPI operations cannot be
optimized by a compiler.  However, it also means that any MPI library
can exploit the newest and best compilers and that the compiler can
be developed without worrying about the impact of MPI on the generated
code---from the compiler's point of view, MPI calls are simply generic
function calls.\footnote{There are some conflicts between the MPI model
and the Fortran language; these are discussed in
\cite[Section 10.2.2]{mpi-forum:mpi2-journal}.  The issues are also
not unique to 
MPI; for example, any asynchronous I/O library faces the same issues
with Fortran.}   The ability of MPI to exploit improvements in other
tools is called \emph{composability}.
Another example is in debuggers; because MPI is simply a library, any
debugger can be used with MPI programs.  Debuggers that are capable of
handling multiple processes, such as TotalView \cite{totalview-web-page}, can
immediately be 
used to debug MPI programs.  Additional refinements, such as an
interface to an abstraction of message passing that is described in 
\cite{pvmmpi99-totalview}, allows users to use the debugger to discover
information about pending communication and unreceived messages.  

More integrated approaches, such as language extensions, have the
obvious benefits, but they also have 
significant costs.  A major cost is the difficulty of exploiting
advances in other tools and of developing and maintaining a large,
integrated system.

OpenMP is an example of a programming model that
achieves the effect of composability with the compilers because OpenMP
requires essentially orthogonal changes to the compiler; that is, most
of the compiler development can ignore the addition of OpenMP in a way
that more integrated languages cannot.

\subsection{Completeness}
\label{sec:completeness}

MPI provides a complete programming model.  Any parallel algorithm can be
implemented with MPI.  Some parallel programming models have sacrified
completeness for simplicity.  For example, a number of programming
models have required that synchronization happens only collectively
for all processes or tasks.  This requirement significantly simplifies the
programming model and allows the use of special hardware affecting all
processes.  Many existing programs also fit into this model; 
data-parallel programs are natural candidates for this model.
But as discussed in Section~\ref{sec:modularity}, many programs
are becoming more complex and are exploiting software components.
Such applications are difficult, if not impossible, to build using restrictive 
programming models. 

Another way to look at this is that while many programs may not be
easy under MPI, no program is impossible.  MPI is sometimes called the
``assembly language'' of parallel programming.  Those making this
statement forget that C and Fortran have also been described as
portable assembly languages.  The generality of the approach should
not be mistaken for an unnecessary complexity.

\subsection{Summary}
Six different requirements have been discussed, along with how MPI
addresses each.  Each of these is \emph{necessary} in a general-purpose
parallel programming system.

Portability and performance are clearly required.  Simplicity and
symmetry cater to the \emph{user} and make it easy to learn and use
safely.  
Composibility is required to prevent the approach from being left
behind by the advance of other tools such as compilers and debuggers.

Modularity, like completeness, is required to ensure that tools can be
built on top of the programming model.  Without modularity, a
programming model is suitable only for turnkey applications.  While
those may be important and easy to identify as customers, they
represent the past rather than the future.  

Completeness, like modularity, is required to ensure that the model supports a large
enough community.  While this does not mean that everyone uses
every function, it means that the functionality that a user may need
is likely to be present.  An early poll of MPI users \cite{osc-poll}
in fact found 
that while no one was using all of the MPI-1 routines, essentially all
MPI-1 routines were in use by someone.  

The open standards process (see
\cite{Hempel99:mpi-standard-process} for a description of the process used to
develop MPI) was an important component in its success.
Similar processes are being adopted by others; see
\cite{kre01:standards} for a description of the
principles and advantages of an open standards process.

\section{Where Next?}

MPI is not perfect.  But any replacement will need to improve on all
that MPI offers, particularly with respect to performance and
modularity, without sacrificing the ability to express any parallel
program.
Three directions are open to investigation: improvements in the MPI
programming model, better MPI implementations, and fundamentally new
approaches to parallel computing.

\subsection{Improving MPI}
\label{sec:improving-mpi}
Where can MPI be improved?  A number of evolutionary
enhancements are possible, many of which can be made by creating tools
that make it easier to build and maintain MPI programs.  

\begin{enumerate}
\item Simpler interfaces.  A compiler (or a preprocessor) could
provide a simpler, integrated syntax.  For example, Fortran 90 array
syntax could be supported without requiring the user to create special
MPI datatypes.  Similarly, the MPI datatype for a C structure could be
created automatically.  Some tools for the latter already exist.
Note that support for array syntax is an example of
support for a subset of the MPI community, many of whom use
data structures that do not map easily onto Fortran 90 arrays.  A
precompiler approach would maintain the composability of the tools,
particularly if debuggers understood preprocessed code.

\item Elimination of function calls.  There is no reason why a
sophisticated system cannot remove the MPI routine calls and replace
them with inline operations, including handling message matching.
Such optimizations have been performed for Linda programs
\cite{lncs589*389} and for MPI subsets \cite{ompi-partialeval}.
Many compilers already perform similar operations for simple numerical
functions like \texttt{abs} and \texttt{sin}.  
This enhancement can be achieved by using preprocessors or precompilers and
thus can maintain the composability of MPI with the best compilers.

\item Additional tools and support for correctness and performance
debugging.  Such tools include editors that can connect send and receive
operations so that both ends of the operation are presented to the
programmer, or performance tools for massively parallel programs.
(Tools such as Vampir and Jumpshot \cite{zaki-lusk-gropp-swider99} are
a good start, but much more can be done to integrate the performance
tool with source-code editors and performance predictors.)

\item Changes to MPI itself, such as read-modify-write additions to
the remote memory access operations in MPI-2.  It turns out to
be surprisingly difficult to implement an atomic fetch-and-increment
operation \cite[Section 6.5.4]{gropp-lusk-thakur:usingmpi2} in MPI-2 using
remote memory 
operations (it is quite easy using threads, but that usually entails a
performance penalty).  

\end{enumerate}

\subsection{Improving MPI Implementations}
Having an implementation of MPI is just the beginning.  Just as the
first compilers stimulated work in creating better compilers by
finding better ways to produce quality code, MPI implementations are
stimulating work on better approaches for implementing the features of
MPI.  
Early work along this line looked at better ways to implement
the MPI datatypes \cite{gropp-swider-lusk99,Traeff:1999:FFE}.  Other
interesting work includes the use of threads to provide a
lightweight MPI implementation \cite{demaine97,PPoPP-99*107}.  This
work is particularly interesting because it involves code
transformations to ensure that the MPI process model is preserved
within a single, multithreaded Unix process.

In fact, several implementations of MPI fail to achieve the available
asymptotic bandwidth or latency.  For example, at least two
implementations from different vendors perform unnecessary copies
(in one case because of layering MPI over a lower-level software that does
not match MPI's message-passing semantics).  These implementations can
be significantly improved.  They also underscore the risk in
evaluating the design of a programming model based on a particular
implementation.

\begin{enumerate}
\item Improvement of the implementation of collective routines
for most platforms. %
One reason, 
ironically, is that the MPI point-to-point communication routines on
which most MPI implementations build their collective routines
are too \emph{high} level.  An alternative approach is to build the
collective routines on top of stream-oriented methods that understand
MPI datatypes.

\item Optimization for new hardware, such as implementations of VIA
or Infiniband. Work in this direction is already taking place,
but more can be done,
particularly for collective (as opposed to point-to-point) communication.

\item Wide area networks (1000 km and more).  In this situation, the
steps used to send a message can be tuned to this high-latency
situation.
In particular, approaches that implement speculative receives
\cite{Tatebe98:mpi-recv-rendezvous}, strategies that make use of
quality of service \cite{rfgkst00:mpichg-qos}, 
or alternatives to IP/TCP may be able to achieve better performance.

\item Scaling to more than 10,000 processes.  Among other things, this
requires better handling of internal buffers; also, some of the routines for
managing process mappings (e.g., \texttt{MPI\_Graph\_create}) do not have
scalable definitions.  

\item Parallel I/O, particularly for clusters.  While parallel file
systems such as PVFS \cite{carns:pvfs} provide support for I/O on
clusters, much more needs to be done, particularly in the areas of
communication aggregation and in reliability in the presence of faults.

\item Fault tolerance. The MPI intercommunicator (providing for
communication between two groups of processes) provides an elegant
mechanism for generalizing the usual ``two party'' approach to fault
tolerance.  Few MPI implementations support fault tolerance in this
situation, and little has been done to develop intercommunicator
collective routines that provide a well-specified behavior in the
presence of faults.

\item Thread-safe and efficient implementations for the support of 
``mixed model'' (message-passing plus threads) programming.  The need
to ensure thread-safety of an MPI implementation used with threads can
significantly increase latency. 
Architecting an MPI implementation to avoid or reduce these penalties
remains a challenge.
\end{enumerate}

\subsection{New Directions}

In addition to improving MPI and enhancing MPI implementations, more
revolutionary efforts should be explored.

One major need is for a 
better match of programming models to the multilevel memory
hierarchies that the speed of light imposes, without adding
unmanageable complexity.  Instead of denying the importance of
hierarchical memory, we need a memory centric view of computing.

MPI's performance comes partly by accident; the two-level memory model
is better than a one-level memory model at allowing the programmer to
work with the system to achieve performance.  But a better approach needs
to be found.

Two branches seem promising.  One is to develop programming models
targeted at hardware similar in organization to what we have today
(see Figure~\ref{fig:generic-parallel}).  The other is to codevelop
both new hardware and new programming models.  For example, hardware
built from processor-in-memory, together with hardware support for rapid
communication of functions might be combined with a programming model
that assumed distributed control.  
The Tera MTA architecture may be a step in such a direction, by
providing extensive hardware support for latency hiding by extensive
use of hardware threads.
In either case, better techniques must be provided for both data
transfer and data synchronization.

Another major need is to make it harder to write incorrect programs.  
A strength of MPI is that
incorrect programs are usually deterministic, simplifying the
debugging process compared to the race conditions that plague
shared-memory programming.  The synchronous send modes (e.g.,  
\texttt{MPI\_Ssend}) may also be used to ensure that a program has no
dependence on message buffering.  

\section{Conclusion}
The lessons from MPI can be summed up as follows: It is more important to make
the hard things possible than it is to make the easy things easy.
Future programming models must concentrate on helping programmers with
what is hard, including the realities of memory hierarchies and the
difficulties in reasoning about concurrent threads of control.

\section*{Acknowledgment}
This
work was supported by the Mathematical, Information, and
Computational Sciences Division subprogram of the Office of
Advanced Scientific Computing Research, U.S.\ Department of Energy,
under Contract W-31-109-Eng-38.

\bibliographystyle{splncs}
\bibliography{papernew}

\end{document}